# Effect of seed layer thickness on Ta crystalline phase and spin Hall angle

K. Sriram[1,#], Jay Pala[1,#], Bibekananda Paikaray[1], Arabinda Haldar[2], Chandrasekhar Murapaka[1*]

[1] Department of Materials Science and Metallurgical Engineering, Indian Institute of Technology, Kandi, Sangareddy, India.
[2] Department of Physics, Indian Institute of Technology, Kandi, Sangareddy, India.

Heavy metal – ferromagnet bilayer structures have attracted great research interest for charge-to-spin interconversion. In this work, we have investigated the effect of the permalloy (Py) seed layer on the Tantalum (Ta) polycrystalline phase and its spin Hall angle. Interestingly, for the same deposition rates the crystalline phase of Ta deposited on Py seed layer strongly depends on the thickness of the seed layer. We have observed a phase transition from α-Ta to (α+β)-Ta while increasing the Py seed layer thickness. The observed phase transition is attributed to the strain at interface between Py and Ta layers. Ferromagnetic resonance-based spin pumping studies reveal that the spin-mixing conductance in the (α+β)-Ta is relatively higher as compared to the α-Ta. Spin Hall angles of α-Ta and (α+β)-Ta are extracted from inverse spin Hall effect (ISHE) measurements. Spin Hall angle of the (α+β)-Ta is estimated to be $\theta_{SH} = -0.15 \pm 0.009$ which is relatively higher than that of α-Ta. Our systematic results connecting the phase of the Ta with seed layer and its effect on the efficiency of spin to charge conversion might resolve ambiguities across various literature and open up new functionalities based on the growth process for the emerging spintronic devices.

1. **INTRODUCTION**

Rapid developments in the field of spin to charge conversion and vice-versa allow electrical control of spin-based phenomena – an essential requirement to integrate spintronic devices with existing microelectronics platform [1–7]. Materials having large spin-orbit coupling (SOC) generates transverse spin current from longitudinal charge current by spin Hall effect (SHE) and interfacial Rashba-Edelstein effect (REE). SHE and REE are responsible for conversion of charge current into a spin current in heavy metals (HM) where spin Hall angle ($\theta_{SH}$) – a parameter that defines the efficiency of charge to spin conversion [8–12]. The converse effect is known as inverse spin Hall effect (ISHE) that converts spin current to charge current is found to be the most promising technique for electrical detection of spin currents [13,14]. One of the key ingredients of the ISHE for converting spin current to a charge current is large SOC and thus, heavy metals such as Ta, W, Pt and Pd are the natural choice [15]. In a typical spin to charge converter, heavy metal (HM)/ferromagnet (FM) interface is used where the spin currents are injected into the HM layer from the FM layer through spin pumping [16–19]. According to Y. Tserkovnyak et al. [20], time-dependent magnetization transfers angular momentum from FM to FM/HM interface via a coupling of the local magnetic moments of the FM to the conduction electrons of the HM. This loss of angular momentum in the Ta layer enhances the ferromagnetic resonance (FMR) linewidth which is an additional damping ($\Delta\alpha$) arising due to the spin pumping to the bulk damping ($\alpha_{FM}$) of FM. In the spin pumping mechanism, oscillating magnetization at the FM/HM interface induces a spin imbalance in the HM layer thereby generating a spin current in the HM. Effective spin-mixing conductance ($g_{\uparrow\downarrow}$) is a key parameter to quantify the efficiency of spin pumping which is a measure of spin current injection from the FM to the adjacent HM sink. In spin to charge conversion, $g_{\uparrow\downarrow}$ is an interfacial parameter, it can be influenced by interfacial texture, morphology and the crystalline phases of HMs [21–27]. Among the choices of HMs, Ta is one of the most studied material in FM/HM system due to the observation of relatively large $g_{\uparrow\downarrow}$ and spin Hall angle ($\theta_{SH}$) [28,29]. Ta is found to possess two different crystalline phases known as stable α-phase and metastable β-phase which are associated with cubic and tetragonal structures, respectively. In the recent past, an extensive research has been carried out on the value of $\theta_{SH}$ for the different phases of Ta and a variety of values have been reported for $\theta_{SH}$ which are as follows: $\theta_{SH}$ = − 0.10 for amorphous Ta [30], $\theta_{SH}$ = − 0.15 for α-Ta [31], $\theta_{SH}$ = − 0.16 for (α+β)-Ta [32], and $\theta_{SH}$ = − 0.10 to − 0.25 for β-Ta [33]. However, the reported large value is in high resistive phase that hinders the realization of low power SOT devices. Therefore, there is an immense need for the deposition the low resistive Ta layers. It is clear that $\theta_{SH}$ strongly depends on the crystalline phase of the Ta which led to several investigations on the structure of Ta interfaced with different FMs [12,34],. However, structural characterizations are focused only on a bare Ta film in most of these studies and the detailed investigations on the effect of a magnetic seed layer on the evolution of Ta polycrystalline phase are elusive. A direct correlation between the crystalline phase of the Ta with the parameters $g_{\uparrow\downarrow}$ and $\theta_{SH}$ is critically important to address a wide variety of the results obtained so far and for the future ISHE-based spintronic devices.

Here, we report a systematic investigation on the dependence of the Ta polycrystalline phase on the ferromagnetic $Ni_{80}Fe_{20}$ or permalloy (Py) seed layer for different Ta growth rates and thicknesses. Detailed structural characterizations have been conducted in order to identify the phase of Ta using grazing incidence X-ray diffraction (GIXRD) technique. We reveal the variation of spin pumping properties i.e., the magnitude of $g_{\uparrow\downarrow}$ as a function of various Ta phases using broadband ferromagnetic resonance (FMR) spectroscopy. Furthermore, we have estimated the spin to charge conversion efficiency parameter, $\theta_{SH}$ for various Ta phases obtained by different Py thickness by using ISHE measurements. Our results correlate critical parameters, $g_{\uparrow\downarrow}$ and $\theta_{SH}$ as a function of the phase of Ta. Our results also open up a potential route for tuning the crystalline phase and SOC of the heavy metal via the seed layer.

2. **EXPERIMENTAL METHODS**

Ta thin films are deposited on naturally oxidized Si <100> substrate by using DC magnetron sputtering technique. Base pressure of the chamber is always kept below 3 ×10$^{-7}$ mbar and deposition pressure ≃5×10$^{-3}$ mbar is maintained during the deposition. Prior to any sample deposition, pre-sputtering of targets was carried out for 2 min with the shutter closed. We have deposited the following series of thin films; series A: Si/Ta(50 & 18 nm) at different deposition rate ($D_R$) = 0.08 – 0.13 nm/s, series B: Si/Py($t_{Py}$)/Ta(18) for $t_{Py}$= 4, 8, 12, 16, and 20 nm, series C: Si/Py($t_{Py}$) at 0.10nm/s, and series D: Si/Py($t_{Py}$=20)/Ta(18) with Ta deposition rate $D_R$ = 0.13 nm/s. Note that we have varied $D_R$ for only Ta and the magnetic layer was always deposited at a fixed deposition rate of $D_R$= 0.10 nm/s in all our samples. The deposition rate of the Ta is tuned by varying the DC power in the range of 60 - 160 watt during the deposition while keeping all other parameters unchanged. Resistivity of Ta and Py is measured by a conventional four-probe method. Structural properties of all the thin films were determined by GIXRD technique using Cu-K$_\alpha$ (λ = 1.5406 Å) radiation source. We have set the incident angle at 1º and performed 2θ scan in the range of 20 – 90º with a scan rate of 0.02º/s. We have employed a lock-in based broadband FMR technique in order to investigate magnetic and magnetization dynamic properties. FMR measurements are carried out in the range of 4 – 16 GHz excitation frequency and 0 to ±300 $mT$ field. ISHE studies are carried out on samples with dimensions of 4 mm × 8 mm. We have measured the voltage drop due to ISHE at the transverse edges of the sample by using Ag paste contacts and all the measurements were carried out at room temperature.

3. **RESULTS AND DISCUSSION**

**3.1. Effect of the deposition rates ($D_R$): Si/Ta (50 & 18 nm)**

In order to understand the effect of the deposition rates on Ta polycrystalline phase, first bare Ta films *i.e.,* series A thin films have been prepared where $D_R$ is varied from 0.08 nm/s to 0.13 nm/s.

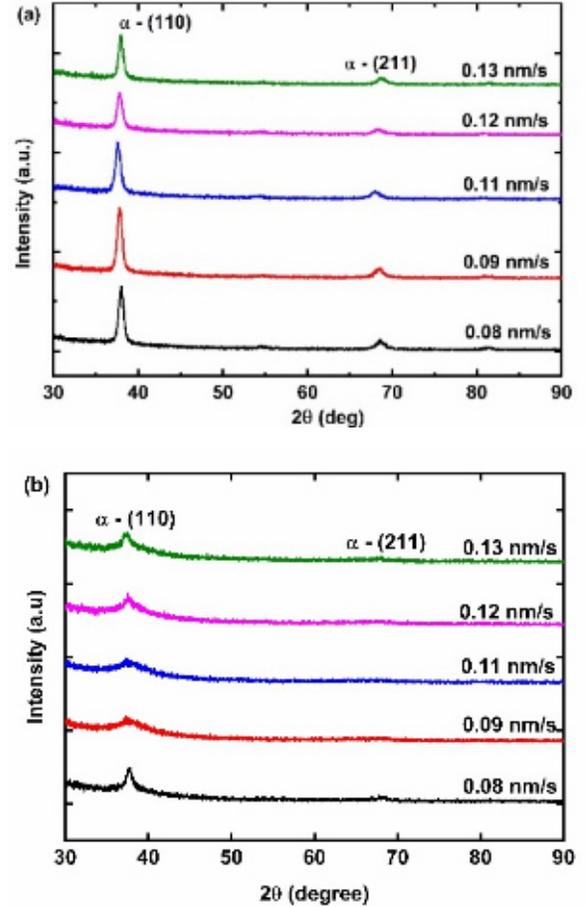

**Fig 1**: GIXRD of Si/Ta($t_{Ta}$). (a) 50nm Ta (b) 18nm Ta grown at different deposition rates from 0.08 to 0.13 nm/s showing the single phase α-Ta.

GIXRD plots of the Si/Ta(50) thin films for different $D_R$ are shown in Fig 1(a). High intensity Bragg diffraction peaks are observed at 2θ ∼ 38º and 2θ ∼ 69.5º and the observed 2θ positions are correspond to (110) and (211) planes of the α-phase of Ta, respectively (denoted by α-Ta) which possesses a body-centered cubic (BCC) crystal structure. The calculated interplanar spacing (*d*) and lattice constant (*a* = *b* = *c*) for the BCC-Ta are 2.37 Å and 3.36 Å, respectively. GIXRD results show that the Ta sputtered directly on the Si substrates exhibit nucleation of single phase α-Ta which is in good agreement with previous reports [35–39]. For the bilayer study, we have taken Ta thickness as 18 nm which is higher than the spin diffusion length ($>2\lambda_{Sd}$) of Ta which also exhibit α-Ta phase deposited on Si. GIXRD results of 18 nm Ta films at different $D_R$ are shown in Fig 1(b). From the GIXRD of Ta at different $D_R$ reveal that at relatively low deposition rates, atoms do have enough relaxation time and diffusivity to occupy energetically favorable atomic positions for equilibrium state which could be the reason for the formation of low energy textured (110) plane. By comparing 50 nm and 18nm thickness of Ta thin films, there is no significant effect on Ta Phase due to its thickness variations.

### 3.2. Effect of seed Py layer: Si/Py($t_{Py}$)/Ta(18); $t_{Py}$ = 4, 8, 12, 16, 20 nm

To investigate the effect of seed Py layer thickness on the Ta polycrystalline phase, we have examined the sample series B. GIXRD results of Si/Py($t_{Py}$)/Ta(18) at $D_R$ = 0.13 nm/s is shown in Fig 2. One can see the stable α-Ta phase of the Ta for $t_{Py}$ = 4 & 8 nm. Interestingly, a phase transition from α-Ta to (α+β)-Ta has been observed for $t_{Py} \geq 12$ nm. In order to understand the effect of Py crystalline nature on Ta, we have deposited series C: Si/Py ($t_{Py}$) to investigate the thickness dependent structural behavior in Py films using GIXRD. GIXRD results are shown in supplementary information S1. GIXRD result shows that prominent (111) plane is observed at 2θ = 44.2° in Py layers corresponding to Face-centered-cubic (FCC) crystal structure. The calculated d-spacing and lattice constant for crystalline phase of Py ($t_{Py}$ = 12, 16 and 20 nm$) are 2.04 Å and 3.54 Å respectively. Interestingly, the observed (111) peak is visible only when the Py thickness is ≥ 12 nm and no prominent diffraction peak is observed below 12 nm thickness. The promotion of (α+β) phase growth of Ta on crystalline Py ($t_{Py} \geq 12$ nm) could be due to the strain at the interface between the crystalline Py and Ta.

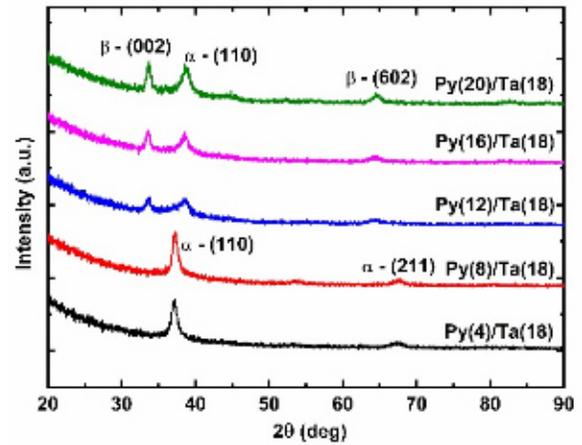

**Fig 2**. GIXRD of Si/Py($t_{Py}$)/Ta(18) bilayers for $t_{Py}$= 4,8,12,16, and 20 nm where Ta deposited at $D_R$= 0.13 nm/s shows Ta phase transition from α-Ta to (α+ β)-Ta phase with a function of seed Py layer thickness.

To further ascertain the phase transition observed in Ta is due to the strain at the interface, we have calculated the lattice parameters of the α-Ta phase from both the α-Ta and (α+β)-Ta deposited on Py of 8 nm and 12 nm thickness, respectively. The lattice constant of α-Ta in Si/Py(12)/Ta(18) is found to be 3.29 Å which is 3.23% smaller than that of lattice constant of the α-Ta in Si/Py(8)/Ta(18). This clearly shows the influence of seed Py layer on Ta crystalline phase. The strain induced at the interface is causing the nucleation of β-Ta along with α-Ta in Si/Py(12)/Ta(18). P. Saravanan et al [40]., have reported that Ta in contact with crystalline Py generates strain due to lattice mismatch between Ta and crystalline Py. This observation is quite evident as well in our Py/Ta system where strain is originated once Py becomes crystallized. The systematic study of A. Fillon et al [41]., on the influence of phase transformation on strain evolution suggests that the strain evolved above critical Py thickness (> 8 nm) due to the increment of lateral volume of sputtered flim after which Py exhibits crystalline nature. From the above-mentioned studies, it can be noted that the crystalline Py thickness is strongly influencing the Ta phase through strain at the Py/Ta interface.

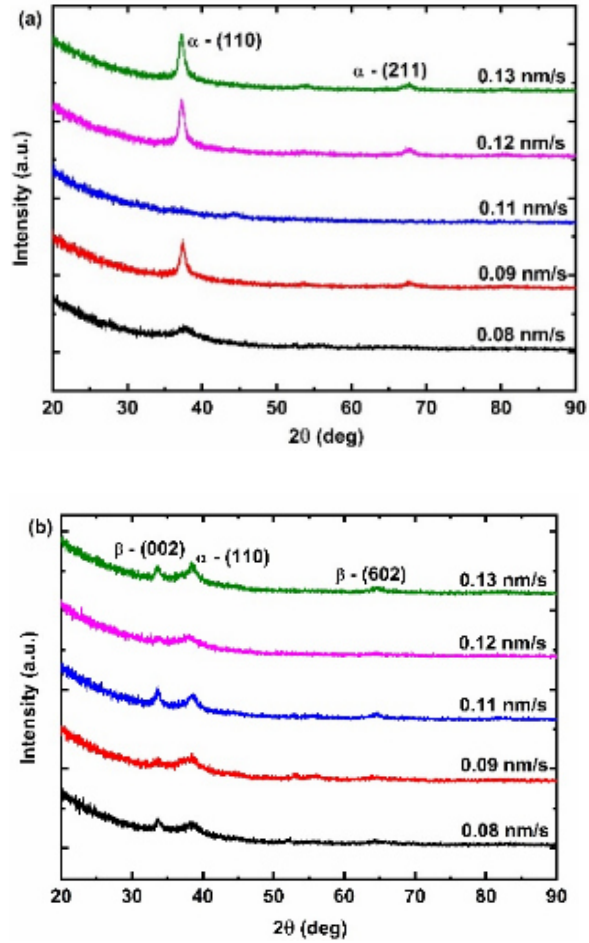

**Fig 3**. (a**)** GIXRD of Si/Py(8)/Ta(18) bilayer showing α-Ta phase for different Ta deposition rate from 0.08 to 0.13 nm/s. (b) GIXRD of Si/Py (12)/Ta (18) bilayer showing (α+ β)-Ta phase for different Ta deposition rate from 0.08 to 0.13 nm/s.

To further confirm the Ta phase transition as a function of the seed Py layer thickness, we have deposited and examined series B at different Ta deposition rates from $D_R$ = 0.08 to 0.13 nm/s. Fig 3. (a & b) shows the GIXRD of Py(8)/Ta(18) and Py(12)/Ta(18) bilayers structures. The Ta deposited on 8 nm Py always exhibits the α-Ta for the deposition rates chosen in our study. In contrast, Ta deposited on 12 nm crystalline Py shows (α+β)-Ta irrespective of the deposition rates. This observation of GIXRD results ascertain that Ta crystalline phase is influenced by seed Py thickness irrespective of the Ta deposition rates considered in this work. It is evident that crystalline seed Py promotes (α+β)-Ta phase by the influence of interfacial strain at Py/Ta interface. From the above discussion, it can be concluded that resultant Ta phase is strongly correlated to the seed Py crystalline nature that depends on Py thickness in our study.

### 3.3. Spin pumping and ISHE for Si/Py($t_{Py}$)/Ta(18); $t_{Py}$= 20 nm

To get an insight on the effect of Ta crystalline phase tuned via seed layer thickness on its spin Hall angle, we have performed spin pumping studies in Si/Py(tpy)/Ta(18) bilayer structures. Therefore, the effect of the Ta phase on spin Hall angle is systematically investigated. First, we have performed FMR measurements on series D [Si/Py($t_{Py}$=20)/Ta(18)]. The derivative of FMR responses shown in Fig 4 (a), are fitted to a derivative Lorentzian function which has symmetric and asymmetric contributions as per the following relation,

$$\frac{dI}{dH} \propto 2K_1 \frac{\Delta H\,(H-H_r)}{[\Delta H^2+(H-H_r)^2]^2} + K_2 \frac{[\Delta H^2-(H-H_r)^2]}{[\Delta H^2+(H-H_r)^2]^2} \quad (1)$$

where *H*, *ΔH*, *H<sub>r</sub>*, *K<sub>1</sub>*, and *K<sub>2</sub>* are the external field, FMR linewidth (full width at half maximum), resonance field, symmetric and asymmetric amplitudes of FMR signal, respectively.

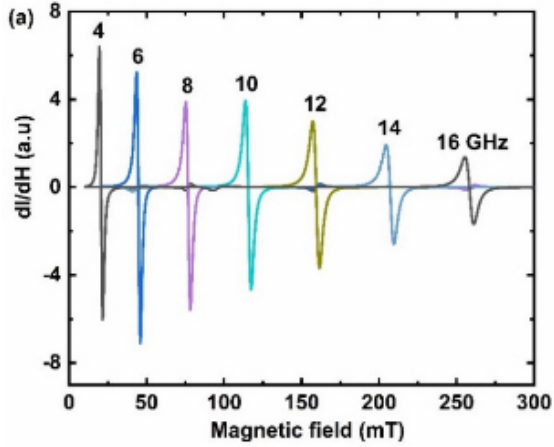

**Fig 4.** Ta deposition rate is set to 0.13 nm/s. (a) Derivative of FMR absorption spectra of Si/Py(20)/Ta(18).

$\Delta H$ and $H_r$ are recorded as fitting parameters from the fit with Equation. (1). Fig 4 (b), shows $H_r$ (*f*) data which are fitted with the Kittel's equation [42]:

$$f = \frac{\gamma}{2\pi}\sqrt{(\mu_0(H_r + H_k)(\mu_0 H_r + \mu_0 H_k + \mu_0 M_{eff})} \quad (2)$$

where $\gamma$, $H_k$, $\mu_0$, $M_{eff}$ are the gyromagnetic ratio ($\gamma$ = 1.85×10$^2$ GHz/T), anisotropy field, vacuum permeability and effective magnetization, respectively. Thus, we obtain $M_{eff}$ and $H_k$ values from the fit with the Kittel equation for all samples. Fig 4 (c), shows the FMR linewidth ($\Delta H$) *vs.* frequency (*f*) plot and the effective Gilbert damping ($\alpha_{eff}$) can be determined from the slope using the following expression:

$$\mu_0 \Delta H = \frac{4\pi \alpha_{eff} f}{\gamma} + \mu_0 \Delta H_0 \quad (3)$$

where $\mu_0 \Delta H_0$ is the inhomogeneous linewidth broadening which is related to the magnetic defects or quality of the film. The fit to the Equation. (3) provides $\alpha_{eff}$ and $\mu_0 \Delta H_0$. The inhomogeneous line width broadening in our samples $\mu_0 \Delta H_0$ is < 1 mT.

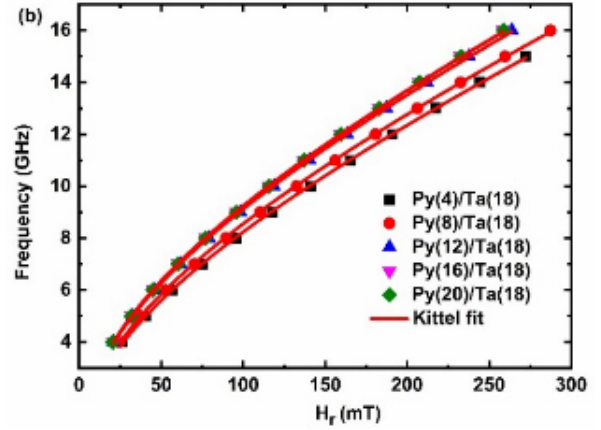

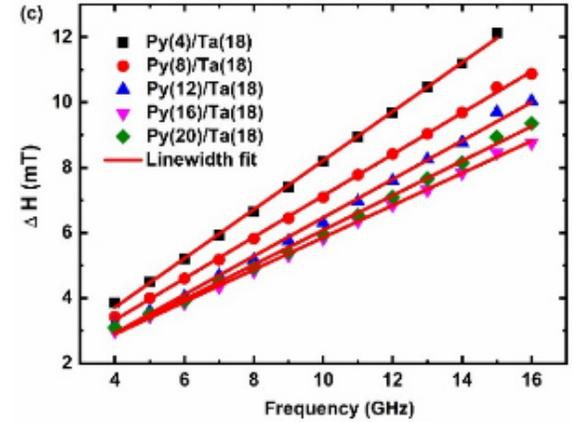

**Fig 4**. Ta deposition rate is set to 0.13 nm/s. (b) Resonance field ($H_r$) vs frequency (*f*) for Si/Py(t$_{Py}$)/Ta(18) where t$_{Py}$ = 4, 8, 12, 16 and 20 nm. (c) FMR linewidth ($\Delta H$) vs frequency (*f*) for Si/Py(t$_{Py}$)/Ta(18) where t$_{Py}$ = 4, 8, 12, 16 and 20 nm.

Additional damping ($\Delta \alpha$) due to spin pumping can be written as,

$$\Delta \alpha = \alpha_{eff(Py/Ta)} - \alpha_{(Py)} \quad (4)$$

where $\alpha_{Py}$ is found to be 0.0066 in our bare Py films. $\alpha_{eff}$ is the direct evidence of spin current induced by spin pumping. The observed $\alpha_{eff}$ for Py/Ta bilayer is more significant when the Py thickness is < 10 nm and decreases at larger Py thickness due to large spin accumulation. It suggests that spin pumping is an interfacial phenomenon and the $\Delta\alpha$ caused by spin pumping is proportional to $1/t_{Py}$ which is consistent with earlier reports [20,32,43–45]. Spin pumping induces non-equilibrium spin accumulation that can be estimated from a parameter called effective spin-mixing conductance ($g_{\uparrow\downarrow}$) by using the following expression,

$$g_{\uparrow\downarrow} = \frac{4\pi M_s t_{FM}}{g\mu_o\mu_B}\left(\alpha_{eff} - \alpha_{Py}\right) \qquad (5)$$

where $g$ (= 2.1) is the spectral splitting constant, $\mu_o$ is the permeability in vacuum, $\mu_B$ is the Bohr magnetron, $M_s$ is the saturation magnetization and $t_{FM}$ is the thickness of Py. We have calculated the spin-mixing conductance for Si/Py($t_{Py}$)/Ta(18) bilayers and the maximum value of $g_{\uparrow\downarrow}$ = 10.1 × 10$^{18}$ m$^{-2}$ is observed for Si/Py(20)/Ta(18) and the minimum value of 7.9 × 10$^{18}$ m$^{-2}$ for Si/Py(8)/Ta(18) which corresponds to (α+β)-phase of Ta and α-phase of Ta, respectively. Spin-mixing conductance ($g_{\uparrow\downarrow}$) is purely an interfacial parameter that quantifies the amount of spin injected from precessing ferromagnet (FM) to the heavy metal (HM) [18,46–48]. E. Simanek et al., [49] and E. Simanek., [50] argued that the enhancement of $g_{\uparrow\downarrow}$ by spin pumping is due to the dynamic electron-electron interaction at FM/HM interface. Theoretical study of magnetization relaxation by Maciej Zwierzycki et al., [45] confirmed that $g_{\uparrow\downarrow}$ is due to coherent scattering of spins within the magnetic exchange length scale. According to Stoner's model discussed by Y. Tserkovnyak et al., [20,51,52] $g_{\uparrow\downarrow}$ can be correlated to the structural property of HM since spin transparency (stoner-enhanced dynamics spin susceptibility) might be dissimilar for different crystal systems. For metallic systems, band structure calculations of the HM give results that are very close to Sharvin conductance (dimensionless conductance means no of transport channels) [53–55]. There could be two reasons for the enhancement of $g_{\uparrow\downarrow}$ in (α+β)-Ta. First reason in our case is the different crystalline phases of the Ta layer that possess different dynamics spin susceptibilities. Due to mixed phase of Ta, there might be more spin injection due to more no of transport channels resulted in large spin-mixing conductance. Second reason is due to the onset of (111) plane in Py (12, 16 and 20 nm) which can potentially change the crystal field effect [56] at Py (12)/Ta (18) and also enhances symmetry breaking [57]. We believe that these are the two major reasons behind the enhancement of the spin-mixing conductance in mixed phase than the single-phase Ta. The variation in the $g_{\uparrow\downarrow}$ is due to the phase fraction of α-Ta and β-Ta. It is worth noting here that the recent works by R. Bansal et al., [23] and A. Kumar et al., [32,58] have also reported similar kind of observations due to the (α+β)-Ta phase. However, theoretical studies on HM structure dependent spin-mixing conductance are elusive and therefore, it will open up a great interest in the community to further dig into using first principle studies. Therefore, it is evident that the interface is better transparent in Py/(α+β)-Ta than in Py/α-Ta for spin injection. In order to investigate the spin-to-charge conversion efficiency, we have considered Si/Py($t_{Py}$)/Ta(18) where $t_{Py}$ = 20 nm, deposited with Ta deposition rate of 0.13 nm/s, respectively. The derivative FMR signal and ISHE voltage for Si/Py(20)/Ta(18) at *f* = 9 GHz, are shown in Fig 5 (a). We have also measured the ISHE voltage for a wide frequency range (4−16 GHz) for Si/Py(20)/Ta(18) with a field sweep of -300 *mT* to +300 *mT* as shown in Fig 5 (b).

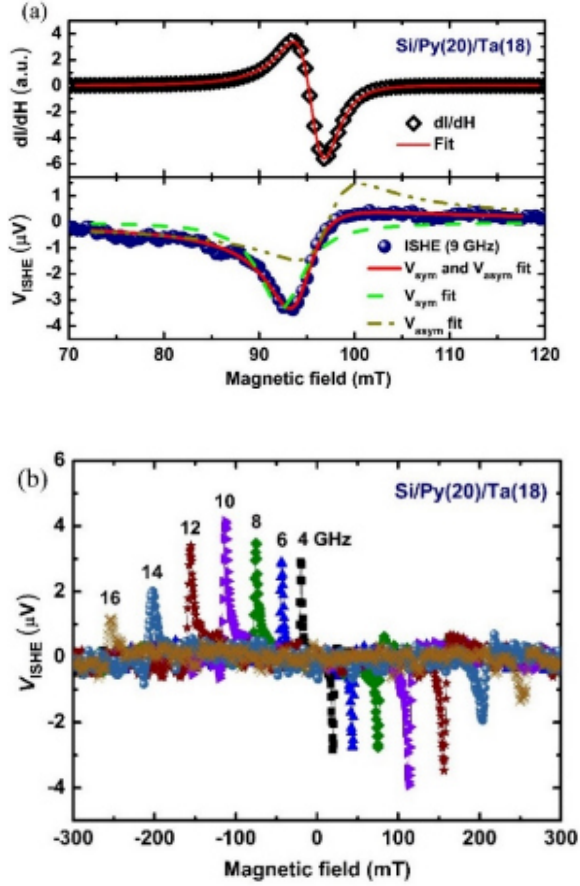

**Fig 5.** (a). Ta deposition rate is set to 0.13 nm/s for Si/Py(20)/Ta(18). Derivative of FMR absorption and corresponding ISHE voltage at 9 GHz excitation frequency with corresponding symmetric and asymmetric fitting. (b). Ta deposition rate is set to 0.13 nm/s for Si/Py(20)/Ta(18). ISHE voltage as function of external field for different GHz frequencies.

In order to disentangle the voltage contribution from ISHE among the all other possible spin rectification effects in our Py/Ta bilayer system, the measured voltage signal is fitted to the Lorentzian equation with a symmetric and an asymmetric contribution [59–61],

$$V = V_{sym} \frac{(\Delta H)^2}{(H-H_r)^2+(\Delta H)^2} V_{asym} \frac{2\Delta H (H-H_r)}{(H-H_r)^2+(\Delta H)^2} \quad (6)$$

where, $V_{sym}$ and $V_{asym}$ are the symmetric and asymmetric components of the measured voltage signal. $V_{sym}$ corresponds to the ISHE and $V_{asym}$ has the contributions from the spin rectification effects such as anisotropic magnetoresistance (AMR) and anomalous Hall Effect (AHE) [62]. The symmetric and asymmetric contribution fit to Eq. (6) is shown in Fig 5 (a). We have found that $V_{sym}$ = 3.6 µV which is dominating as compared to $V_{asym}$ = 1.3 µV. Therefore, the major contribution of the observed voltage signal can be attributed to the ISHE. The spin-pumping induced ISHE in our system enables the estimation of spin current density ($J_S$) and $\theta_{SH}$ of the Ta layer. In Py/Ta film, the magnitude of the spin current injected from the Py to the Ta layer can be evaluated from spin current density formulation [20], which can be expressed as,

$$|J_S| = \left(\frac{g_{\uparrow\downarrow}\hbar}{8\pi}\right)\left(\frac{\mu_0 h_{rf}\gamma}{\alpha_{eff}}\right)^2 \left[\frac{\mu_0 M_S\gamma+\sqrt{(\mu_0 M_S\gamma)^2+16(\pi f)^2}}{(\mu_0 M_S\gamma)^2+16(\pi f)^2}\right]\left(\frac{2e}{\hbar}\right) \quad (7)$$

where $\mu_0 h_{rf}$ is the rf magnetic field which is 0.06 mT in our measurements. The spin current density described in Eq. (7) is converted into an electromotive force $V_{ISHE}$ due to the ISHE in the Ta layer induced by the spin pumping as per the following relation [63],

$$V = \left(\frac{1}{\frac{t_{Py}}{\rho_{Py}}+\frac{t_{Ta}}{\rho_{Ta}}}\right) w\, \theta_{SH}\, \lambda_{sd}\, \tanh\left(\frac{t_{Ta}}{2\lambda_{sd}}\right) |J_S| \quad (8)$$

Where $\rho_{Py}$ and $\rho_{Ta}$ are the resistivities of Py and Ta thin films, respectively [34,50]. The parameters w, $t_{Ta}$, $t_{Py}$ are the width of the signal line (= 200 µm), thickness of the Ta and Py layer, respectively. The spin diffusion length ($\lambda_{sd}$) is considered to be 3 nm for Ta [64]. Interestingly, we have observed that Ta thin film with (α+β)-Ta phase shows a higher value of spin Hall angle of −0.15 ± 0.009 than α-phase Ta whose $\theta_{SH}$ value is = −0.10 ± 0.008. From the observed $\theta_{SH}$ values, it is evident that is spin to charge conversion efficiency is directly correlated with phase of Ta and the estimated $\theta_{SH}$ values are good agreement with reported values [32]. The enhanced spin Hall angle in the Py/Ta has low longitudinal resistance due to the presence of mixed phase where the major contribution comes from extrinsic mechanism such as skew scattering and side jump scattering as reported by Kumar. A et al., [32], [58], [23]. The crystalline Py can enhance the $\theta_{SH}$ by enlarging interfacial symmetry breaking at Py/Ta interface. Therefore, the key reason for the relatively large $\theta_{SH}$ observed in the Py/(α+β)-Ta phase is due to the combined effect of low longitudinal resistance and the enhanced interfacial symmetry breaking. This work presents a promising method for the engineering the crystalline phase of Ta via seed Py thickness and crystallinity which in turn assist in tuning spin Hall angle. It also reveals that the effect of thickness and crystalline nature of the seed ferromagnetic (FM) layer on the crystalline phase of Ta cannot be ignored. It shows that Ta films deposited on bare Si substrate and FM seed layer can exhibit different crystalline phase hence exhibit different spin-charge conversion efficiency. Our systematic investigation on Py/Ta may provide a viable and alternative way to tune the spin conversion efficiency via seed layer crystallinity and thickness. Moreover, this study improves the understanding on the seed layer influence on HMs phase transition and effect of stack configuration on the performance of SOT based devices.

## 4. SUMMARY

The effect of the magnetic seed layer on the phase of Ta has been investigated in detail in the Py/Ta heterostructures. First, the phase of bare Ta films ($t_{Ta}$= 18, 50 nm) on Si-substrate has been characterized as a function of deposition rate showing α-Ta phase for $D_R$ < 0.2 nm/s and (α+β)-Ta phase beyond it. The phase of the Ta is then systematically studied by depositing it (at $D_R$ = 0.13 nm/s) on different thicknesses Py films which were sputtered on Si-substrate. Si/Py($t_{Py}$)/Ta(18) bilayers reveal α-Ta phase for $t_{Py}$= 4, 8 nm and (α+β)-Ta for $t_{Py} \geq 12$ nm which is critical thickness. Thus, an onset of tetragonal structure associated with the β-Ta phase has been shown in addition to the α-Ta phase with the increasing thickness of the magnetic seed layer. Usually, the thickness of Py is varied in Py/Ta heterostructures for the investigation of ISHE and hence the phase of Ta plays an important role for different important parameters like $g_{\uparrow\downarrow}$ and $\theta_{SH}$. An enhanced spin pumping of $g_{\uparrow\downarrow} = 10.1 \times 10^{18}\ m^2$ is observed in Si/Py($t_{Py}$)/Ta(18) for the (α+β)-Ta phase ($t_{Py} \geq 12$ nm) as compared to $g_{\uparrow\downarrow} = 7.9 \times 10^{18}\ m^2$ for the α-Ta ($t_{Py} < 12$ nm) by using FMR measurements. Consequently, the spin-to-charge conversion efficiency is found to be higher for the (α+β)-Ta phase ($\theta_{SH} = -0.15 \pm 0.009$) than the α-Ta phase ($\theta_{SH} = -0.10 \pm 0.008$) by performing ISHE measurements. Our results demonstrate a strong correlation between the phase of Ta and the observed spin-to-charge conversion parameters in Py/Ta heterostructures. Therefore, this work has potential implications in designing efficient ISHE-based spintronic devices via seed layer thickness.

## AUTHOR CONTRIBUTIONS

CM and AH conceived the idea and coordinated the project. KS and JP have prepared the samples. KS has performed the structural characterization and analysed the XRD data. BP and JP have performed the FMR based spin pumping and ISHE experiments. KS, BP and JP have analysed the FMR and ISHE data. Manuscript is written by KS, AH and CM.

**CONFLICTS OF INTEREST**

The authors declare no conflict of interest

**ACKNOWLEDGEMENT**

CM would like to acknowledge funding from SERB-Early Career Research Award (ECR/2018/002664). AH would like to acknowledge funding from Ramanujan Fellowship (SB/S2/RJN-118/2016), Department of science and Technology, India. KS would like to acknowledge the fellowship from the SERB project (ECR/2018/002664). BP would like to acknowledge fellowship from the Department of science and Technology, India (DST/INSPIRE Fellowship/ [IF180927]). The authors would like to thank Dr. Gajendranath Choudray from Electrical Engineering Department, IIT Hyderabad for giving access to nanovoltmeter.

[#]KS and JP have equally contributed to this work.